\documentclass{article}

\usepackage{arxiv}

\usepackage[utf8]{inputenc} 
\usepackage[T1]{fontenc}    
\usepackage{hyperref}       
\usepackage{url}            
\usepackage{booktabs}       
\usepackage{amsfonts}       
\usepackage{nicefrac}       
\usepackage{microtype}      
\usepackage{graphicx}
\usepackage{doi}

\usepackage{bm} 
\usepackage{amsbsy} 
\usepackage{float}
\usepackage{colordvi,color}
\usepackage[dvipsnames]{xcolor}  

\usepackage{threeparttable} 

\usepackage{xspace}

\usepackage{mathtools}
\usepackage{array}
\newcolumntype{C}[1]{>{\centering\arraybackslash}p{#1}}
\newcolumntype{R}[1]{>{\raggedleft\arraybackslash}p{#1}}

\title{Reduced SIGMA Basis Sets: a new family of SIGMA basis sets for molecular calculations}

\author{ \href{https://orcid.org/0000-0002-3003-213X}{\includegraphics[scale=0.6]{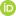}\hspace{1mm}Ignacio Ema}\thanks{
        } \\
    Departamento de Química Física Aplicada, \\
    Facultad de Ciencias, \\
    Universidad Autónoma de Madrid, \\
    Spain \\
    \texttt{nacho.ema@uam.es} \\
    \And
    \href{https://orcid.org/0000-0002-3824-8277}{\includegraphics[scale=0.6]{orcid.png}\hspace{1mm}
    Jesús San Fabián} \\
    Departamento de Química Física Aplicada, \\
    Facultad de Ciencias, \\
    Universidad Autónoma de Madrid, \\
    Spain \\
    \texttt{jesus.sanfabian@uam.es} \\
    \And
    \href{https://orcid.org/0000-0002-6957-1050}{\includegraphics[scale=0.6]{orcid.png}\hspace{1mm}Guillermo Ramírez} \\
    Departamento de Química Física Aplicada, \\
    Facultad de Ciencias, \\
    Universidad Autónoma de Madrid, \\
    Spain \\
    \texttt{guillermo.ramirez@uam.es} \\
    \And
    \href{https://orcid.org/0000-0002-9676-7534}{\includegraphics[scale=0.6]{orcid.png}\hspace{1mm}Rafael López} \\
    Departamento de Química Física Aplicada, \\
    Facultad de Ciencias, \\
    Universidad Autónoma de Madrid, \\
    Spain \\
    \texttt{rafael.lopez@uam.es} \\
    \And
    \href{https://orcid.org/0000-0002-1940-422X}{\includegraphics[scale=0.6]{orcid.png}\hspace{1mm}José Manuel García de la Vega} \\
    Departamento de Química Física Aplicada, \\
    Facultad de Ciencias, \\
    Universidad Autónoma de Madrid, \\
    Spain \\
    \texttt{garcia.delavega@uam.es} \\
}

\renewcommand{\shorttitle}{\textit{arXiv} Reduced sigma basis sets}
\hypersetup{
pdftitle={A template for the arxiv style},
pdfsubject={q-bio.NC, q-bio.QM},
pdfauthor={David S.~Hippocampus, Elias D.~Striatum},
pdfkeywords={Gaussian functions, Basis sets, Atomic energies, Molecular calculations},
}

\begin{document}

\newcommand{\be}{\begin{equation}}
\newcommand{\ee}{\end{equation}}
\newcommand{\ba}{\begin{array}}
\newcommand{\ea}{\end{array}}
\newcommand{\baa}{\begin{eqnarray}}
\newcommand{\eaa}{\end{eqnarray}}
\newcommand{\Dcal}{{\mathcal{D}}}
\newcommand{\Mcal}{{\mathcal{M}}}
\newcommand{\Ncal}{{\mathcal{N}}}
\newcommand{\Rcal}{\mathcal{R}}  
\newcommand{\Vcal}{\mathcal{V}}
\newcommand{\Ane}{\mathbf{A}}
\newcommand{\Bne}{\mathbf{B}}
\newcommand{\Cne}{\mathbf{C}}
\newcommand{\Dne}{\mathbf{D}}
\newcommand{\Ene}{{\bf E}}
\newcommand{\Fne}{{\bf F}}
\newcommand{\kne}{\mathbf{k}}
\newcommand{\Mne}{{\bf M}}
\newcommand{\Pne}{{\bf P}}
\newcommand{\rne}{{\bf r}}
\newcommand{\Rne}{{\bf R}}
\newcommand{\Sne}{{\bf S}}
\newcommand{\Gtilde}{{\widetilde{G}}}
\newcommand{\arsBS}{$a\sigma$BS0\xspace}
\newcommand{\arsDZ}{a$\sigma$DZ0\xspace}
\newcommand{\arsTZ}{a$\sigma$TZ0\xspace}
\newcommand{\arsQZ}{a$\sigma$QZ0\xspace}
\newcommand{\asBS}{$a\sigma$BS\xspace}
\newcommand{\asDZ}{a$\sigma$DZ\xspace}
\newcommand{\asTZ}{a$\sigma$TZ\xspace}
\newcommand{\asQZ}{a$\sigma$QZ\xspace}
\newcommand{\asXZ}{a$\sigma$XZ\xspace}
\newcommand{\arsXZ}{a$\sigma$XZ0\xspace}
\newcommand{\sBS}{$\sigma$BS\xspace}
\newcommand{\rsBS}{$\sigma$BS0\xspace}
\newcommand{\sBSdos}{$\sigma$BS2\xspace}
\newcommand{\sDZ}{$\sigma$DZ\xspace}
\newcommand{\sTZ}{$\sigma$TZ\xspace}
\newcommand{\sQZ}{$\sigma$QZ\xspace}
\newcommand{\sXZ}{$\sigma$XZ\xspace}
\newcommand{\rsXZ}{$\sigma$XZ0\xspace}
\newcommand{\rsTZ}{$\sigma$TZ\xspace}
\newcommand{\rsQZ}{$\sigma$QZ\xspace}
\newcommand{\rsDZ}{$\sigma$DZ\xspace}
\newcommand{\azul}[1]{{\color{blue}{#1}}}
\newcommand{\verde}[1]{{\color{OliveGreen}{#1}}}
\newcommand{\rojo}[1]{{\color{red}{#1}}}
\newcommand{\C}{\mathrm{C}}
\renewcommand{\H}{\mathrm{H}}
\newcommand{\N}{\mathrm{N}}
\renewcommand{\O}{\mathrm{O}}
\renewcommand{\S}{\mathrm{S}}

\def\arraystretch{1.5} 

\maketitle

\begin{abstract}
    A new family of Gaussian-type SIGMA basis sets, termed reduced SIGMA basis sets, 
    is introduced and preliminarily tested. Sharing the same composition as Dunning 
    basis sets, they enhance performance by reducing linear dependencies in large 
    systems, thereby improving convergence and lowering computational costs for such 
    systems..
\end{abstract}


\section{\label{sec:1}Introduction}
New basis sets (BS) consisting of Gaussian-type orbitals (GTO), derived 
from the previously reported SIGMA BS (\sBS) \cite{Ema2023,Ema2025}, is presented. 
The new BS are named reduced SIGMA (\rsBS) and have the same composition per shell as 
Dunning correlation-consistent BS \cite{Dunning1989,Woon1993}, ranging from DZ to QZ, 
both standard and augmented (\arsBS). The composition of the (a)\rsBS is summarized in Table~\ref{tab:1}.

In particular, the new primitives added in the augmented version (\arsBS) have 
exponents higher than those present in Dunning BS of the same composition, without 
compromising the description of properties related to the diffuse part of the 
electronic cloud. Furthermore, the diffuse primitives of the $s$ and $p$ shells are 
not directly added to the BS but are incorporated through general contractions. 
Together, these features make the \arsBS less susceptible to linear quasi-
dependencies than their Dunning counterparts, resulting in improved convergence in 
terms of both accuracy and computational efficiency for large molecular systems. 
Standard basis sets show equivalent performance between \rsBS and Dunning sets and 
are included to provide a foundation for the augmented versions and for completeness.

Unlike the $\sigma$BS, the requirement that all primitives with the same exponent and 
$l < L$ must also be present in the set is removed in the \rsBS, and the number of 
primitives is reduced compared to the former. This slightly diminishes the quality of 
the results relative to the $\sigma$BS (which are designed for high-precision 
calculations), but achieves a good compromise between accuracy and computational 
cost. The \rsBS are particularly competitive when applied to large systems with 
augmented basis sets.

This work is organized as follows. In Section \ref{sec:2}, the procedure for 
developing SIGMA BS is outlined, and the contraction scheme and composition are 
described. In Section \ref{sec:3}, some preliminary testing on molecular calculations 
is reported.

\section{\label{sec:2}Development of reduced SIGMA basis sets}

Normalized spherical Gaussian primitive functions (pGTO), $g_{lm}$, are defined 
as:
  
\be \label{eq:1.1}
g_{lm}(\xi,\rne) = \Ncal_{l,\xi}^{r} \; \Ncal_{lm}^\Omega \; 
e^{-\xi \, r^2} \; z_l^m(\rne)
\ee
where $z_l^m (\rne)$ are unnormalized regular solid harmonics:  
  
\be   \label{eq:1.2}  
z_l  (\rne) = r^l \; (-1)   \; P_l^{|m|}(\cos \theta)  
\;   
\left\{  
\begin{array}{lr}  
\cos m \phi & \hspace*{5mm} (m \ge 0) \\  
\sin |m| \phi & \hspace*{5mm}  (m <0)  
\end{array}  
\right.  
\ee  
$P_l^m$ being the associate Legendre functions (see ~\cite{Gradshteyn} Eq 8.751.1), and $\Ncal_{\xi_i}$ and $\Ncal_{LM}^\Omega$, the 
radial and angular normalization constants given by:

\be \label{eq:1.3}
\Ncal_{l,\xi}^{r} = 2^{l+1} \; \sqrt{\frac{\xi^l}{(2l+1)!!}} \; \; 
\left[\frac{(2 \, \xi)^3}{\pi}\right]^{1/4} 
\ee

\be \label{eq:1.4}
\Ncal_{lm}^\Omega = \left[\frac{(2l+1) \; (l-|m|)!}{2 \, (1+\delta_{m0}) \;
\pi \; (l+|m|)!}\right]^{1/2}
\ee

Usual GTO BS consist of contractions 
of Gaussian primitives. They are known as contracted Gaussian functions (CGTO), 
$G_{lm}$, 
defined as:

\be \label{eq:1.5}
G_{lm}(\rne) = \sum_{i=1}^{N_G} c_i \; g_{lm,i}(\rne)
\ee
where $i$ labels the primitives in the set, and the coefficients $c_i$ 
are determined by different procedures 
subject to the 
normalization of the GTO. The number of coefficients
and their values depend on the recipe used in the construction of a particular basis set.

To simplify the notation, (a)XZ will be used below as an abbreviation for Dunning 
(aug-)cc-pVXZ  families. The equivalent \rsBS will be denoted as (a)\rsXZ. 
In each of the SIGMA families, basis sets ranging from DZ to QZ
have been considered.

\renewcommand{\arraystretch}{1.2}

\begin{table}[H]
\caption{Composition of SIGMA basis sets for atoms H to Ar.
\label{tab:1}}
\begin{center}
\begin{tabular}{|l|rl|rl|}
\hline
BS  & $\#$ & Primitives & $\#$  & Contractions \\
\hline
 & \multicolumn{4}{c|}{First row: H and He.} \\
\hline
 $\sigma$DZ0    &   8 & ( 5s,  1p )               &  5 & [ 2s,  1p ]             \\
a$\sigma$DZ0    &  12 & ( 6s,  2p )               &  9 & [ 3s,  2p ]             \\
 $\sigma$TZ0    &  17 & ( 6s,  2p, 1d )           & 14 & [ 3s,  2p, 1d ]         \\
a$\sigma$TZ0    &  26 & ( 7s,  3p, 2d )           & 23 & [ 4s,  3p, 2d ]         \\
 $\sigma$QZ0    &  33 & ( 7s,  3p, 2d, 1f )       & 30 & [ 4s,  3p, 2d, 1f ]     \\
a$\sigma$QZ0    &  49 & ( 8s,  4p, 3d, 2f )       & 46 & [ 5s,  4p, 3d, 2f ]     \\
\hline
 & \multicolumn{4}{c|}{Second row: Li to  Ne.} \\
\hline
 $\sigma$DZ0  &  29 & ( 9s,  5p, 1d)             &  14    &[ 3s, 2p, 1d ]             \\
a$\sigma$DZ0  &  38 & (10s,  6p, 2d)             &  23    &[ 4s, 3p, 2d ]             \\
 $\sigma$TZ0  &  45 & (10s,  6p, 2d, 1f)         &  30    &[ 4s, 3p, 2d, 1f ]         \\
a$\sigma$TZ0  &  61 & (11s,  7p, 3d, 2f)         &  46    &[ 5s, 4p, 3d, 2f ]         \\
 $\sigma$QZ0  &  71 & (12s,  7p, 3d, 2f, 1g)     &  55    &[ 5s, 4p, 3d, 2f, 1g ]     \\
a$\sigma$QZ0  &  96 & (13s,  8p, 4d, 3f, 2g)     &  80    &[ 6s, 5p, 4d, 3f, 2g ]     \\
\hline
 & \multicolumn{4}{c|}{Third row: Na to Ar.} \\
\hline
 $\sigma$DZ0    &  44 & (12s,  9p, 1d)             &  18    &[ 4s, 3p, 1d ]             \\
a$\sigma$DZ0    &  53 & (13s, 10p, 2d)             &  27    &[ 5s, 4p, 2d ]             \\
 $\sigma$TZ0    &  62 & (15s, 10p, 2d, 1f)         &  34    &[ 5s, 4p, 2d, 1f ]         \\
a$\sigma$TZ0    &  78 & (16s, 11p, 3d, 2f)         &  50    &[ 6s, 5p, 3d, 2f ]         \\
 $\sigma$QZ0    &  90 & (16s, 12p, 3d, 2f, 1g)     &  69    &[ 6s, 5p, 3d, 2f, 1g ]     \\
a$\sigma$QZ0    & 115 & (17s, 13p, 4d, 3f, 2g)     &  84    &[ 7s, 6p, 4d, 3f, 2g ]     \\
\hline
\end{tabular}
\end{center}
\end{table}

The computation of the primitive exponents and contraction coefficients follows the same guidelines previously described\cite{Ema2023,Ema2025}, while eliminating the requirement that all primitives with the same exponent and $l < L$ be present in the set.

All the calculations reported have been carried out using MOLPRO\cite{MOLPRO} 
in a system with an Intel(R) Xeon(R) Platinum 8362 CPU @ 2.80GHz processor.

\section{\label{sec:3}Results}

To test the performance of the (a)\rsXZ, calculations have been carried out on several 
systems, comparing the results with those obtained using the equivalent Dunning basis sets.

\begin{figure}[H]
    \centering
    \includegraphics[width=0.48\textwidth]{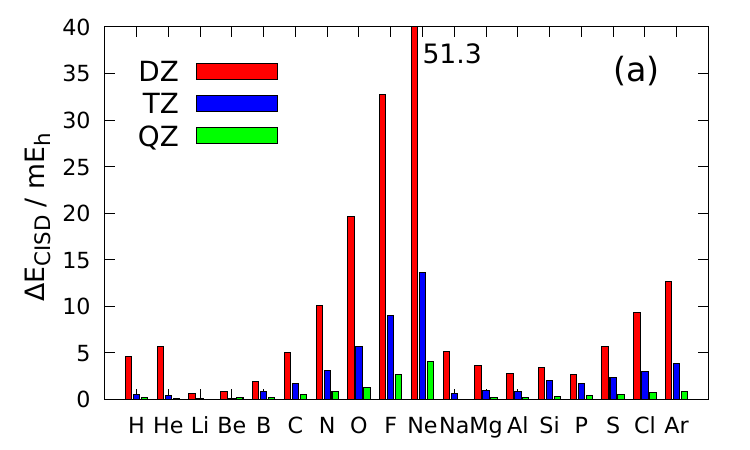}
    \includegraphics[width=0.48\textwidth]{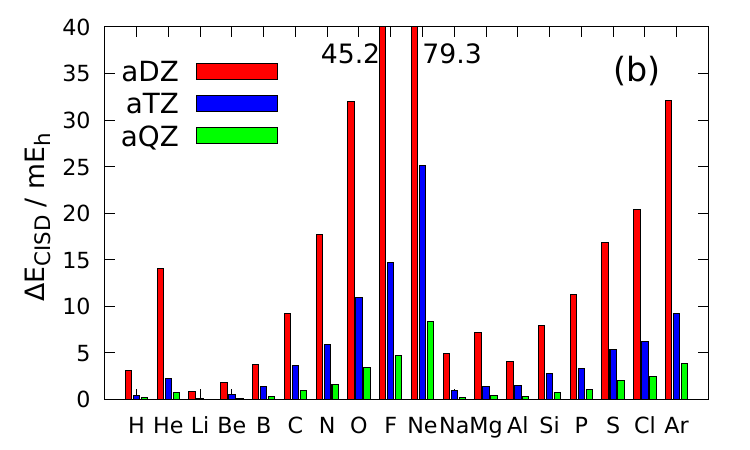}
    \caption{\label{fig:1}CISD energy differences, $\Delta E = E_{Dunning}-E_{SIGMA}$, for standard and augmented BS from double to quadruple size in mE$_h$.}
\end{figure}


In atomic calculations, the energies obtained with the (a)\rsXZ are lower than those of 
the Dunning (a)XZ in all cases, at both the HF and CISD levels.
In Fig.\ref{fig:1}, the differences in CISD energies, computed as
$\Delta E = E_{\text{Dunning}} - E_{\text{SIGMA}}$,
are plotted for the standard (Fig.\ref{fig:1}a) and augmented (Fig.~\ref{fig:1}b) BS.
As can be seen, the differences across a period increase with the atomic number and 
become smaller as the size of the BS increases.
The smallest differences are observed for the alkali metals, while the largest appear 
for the noble gases. As expected, these differences decrease as the quality of the basis 
set increases.
The differences are always positive, confirming that, from a variational point of view,
(a)\rsXZ outperform the (a)XZ. This pattern is observed for both the standard and 
augmented versions.

The performance of (a)\rsXZ in high-quality calculations of small systems has also been analyzed by computing the reaction energy for carbamic dipeptide formation at the CCSD computational level. The results are summarized in Table~\ref{tab:2}, where the a5Z is taken as a reference.
As can be seen, the absolute energies for the reactant (carbamic acid, 
$\text{CH}_3\text{NO}_2$) and products (carbamic dipeptide, 
$\text{C}_2\text{H}_4\text{N}_2\text{O}_3$, and water) are lower with (a)\rsXZ than with 
(a)XZ, and the reaction energies are similar and close to the reference, except in the 
aDZ and \rsDZ cases.
Notably, the transition from standard to augmented BS does not bring the aXZ results 
closer to the reference, unlike the shift from \rsXZ to \arsXZ.

In addition, the average differences in the optimized geometry distances of the 
dipeptide with respect to the reference are given (as percentages) in the last column of 
Table~\ref{tab:2}, considering only distances smaller than 3~\AA.
It can be observed that the geometries computed with \arsXZ are closer to the reference 
than those computed with aXZ, and again, no improvement is seen when moving from XZ to 
aXZ, in contrast to the improvement found when transitioning from \rsXZ to \arsXZ.

To illustrate the performance at other computational levels, B3LYP-D3, HF, and MP2 
calculations of the isomerization between anthracene and phenanthrene have been 
performed using \arsXZ and aXZ basis sets.
In Table~\ref{tab:3}, total energies for the ground states of both molecules are 
reported, along with the corresponding isomerization energy.
Once again, the absolute energies are lower with \arsXZ than with aXZ, although the 
differences in the computed isomerization energies are less pronounced.
The computational time is also provided.
As can be seen, the computational cost is slightly lower (approximately 10–20\%) with 
\arsXZ than with aXZ.

Finally, to test the numerical stability of the basis sets, a calculation on emeraldine 
(C$_{48}$H$_{38}$N$_8$)\cite{emeraldine} was carried out with \arsDZ and aDZ.
The \arsDZ set (2584 primitives and 1630 contractions) yielded an energy of 
$-2252.0008947$ E$_h$, requiring 5.3 hours of computational time.
In contrast, the calculation with the aDZ basis set (2378 primitives, 1630 contractions) 
failed to converge after more than ten hours.
An analysis of the overlap matrix eigenvalues showed that the lowest eigenvalue with 
\arsDZ ($1.0 \cdot 10^{-6}$) was over three orders of magnitude higher than that of aDZ 
($3.2 \cdot 10^{-10}$), which may account for the convergence issues observed.


\begingroup

\renewcommand{\arraystretch}{1.5}

\begin{table}[H]
\caption{\label{tab:2}
CCSD calculations for the carbamide dipeptide formation.
Reaction energies in kJ/mol. Average relative error in distances in the dipeptide for distances smaller 
than 3~\AA, reference: Dunning 5Z.
}
\begin{center}    
\begin{tabular}{|c|ccc|c|c|c|}
\hline
& \multicolumn{5}{c|}{$\C\H_3\N\O_2 + \C\H_3\N\O_2 \rightarrow \C_2\H_4\N_2\O_3 + \H_2\O$} \\
\hline
BS  &  $\C\H_3\N\O_2$ & $\H_2\O$ & $\C_2\H_4\N_2\O_3$ &  $\Delta$E & \%dis.Av.err. \\
\hline 
DZ  & $-244.52569154$ & $-76.23820611$ & $-412.79467452$ & 48.58 & 0.849 \\ 
TZ  & $-244.76550872$ & $-76.32455655$ & $-413.18798966$ & 48.50 & 0.198 \\
QZ  & $-244.84002164$ & $-76.35081232$ & $-413.31143064$ & 46.73 & 0.038 \\ 
5Z  & $-244.86399283$ & $-76.35952965$ & $-413.35090852$ & 46.07 & reference \\
 \hline
aDZ & $-244.58547699$ & $-76.26863314$ & $-412.88754522$ & 38.79 & 1.044 \\ 
aTZ & $-244.78528577$ & $-76.33367020$ & $-413.22033227$ & 43.50 & 0.250 \\
aQZ & $-244.84757214$ & $-76.35421652$ & $-413.32384035$ & 44.86 & 0.066 \\ 
 \hline
$\sigma$DZ0 & $-244.57358696$ & $-76.25944839$ & $-412.8685733$2 & 50.28 & 0.805 \\
$\sigma$TZ0 & $-244.77787669$ & $-76.32964137$ & $-413.2078367$9 & 47.98 & 0.212 \\
$\sigma$QZ0 & $-244.84293707$ & $-76.35168122$ & $-413.3161471$0 & 47.38 & 0.053 \\
 \hline
a$\sigma$DZ0 & $-244.68208124$ & $-76.30112309$ & $-413.04605304$ & 44.60 & 0.549 \\
a$\sigma$TZ0 & $-244.81668850$ & $-76.34481577$ & $-413.27175060$ & 44.14 & 0.127 \\
a$\sigma$QZ0 & $-244.85637681$ & $-76.35732145$ & $-413.33818247$ & 45.29 & 0.030 \\
\hline 
\end{tabular}
\end{center}
\end{table}

\endgroup


\begin{table}[H]   
\caption{\label{tab:3}
B3LYP-D3, HF and MP2 calculations for ground states of anthracene 
and phenanthrene, and energy for Anthracene $\rightarrow$ Phenathrene isomerization 
reaction.  Total energies in E$_h$. Reaction energy in kJ/mol. Computational time in 
hours}
\begin{center}    
\setlength{\tabcolsep}{5pt}
\begin{tabular}{|c|cc|cc|c|c|}
\hline
BS  & \#prim. & \#cont. & Anthracene & Phenantrene & $\Delta$E & Time \\
\hline
\multicolumn{7}{|c|}{B3LYP-D3} \\
\hline
aDZ &  600 &  412 & -539.250407 & -539.260365 & -26.14 &   0.6 \\
aTZ & 1062 &  874 & -539.381024 & -539.390594 & -25.13 &   5.0 \\
aQZ & 1728 & 1580 & -539.416445 & -539.425971 & -25.01 &    52 \\
 \hline
a$\sigma$DZ0 &  652  &  412 & -539.343246 & -539.352779 & -25.03 &   0.5 \\
a$\sigma$TZ0 & 1114  &  874 & -539.408244 & -539.417795 & -25.07 &   4.6 \\
a$\sigma$QZ0 & 1834  & 1580 & -539.425093 & -539.434593 & -24.94 &    48 \\
\hline
\multicolumn{7}{|c|}{HF} \\
\hline
aDZ &  600 &  412 & -536.040994 & -536.054752 & -36.12 &   0.6 \\
aTZ & 1062 &  874 & -536.156692 & -536.170200 & -35.46 &   4.7 \\
aQZ & 1728 & 1580 & -536.186136 & -536.199584 & -35.31 &    49 \\
 \hline
a$\sigma$DZ0 &  652  &  412 & -536.127751 & -536.141228 & -35.38 &   0.5 \\
a$\sigma$TZ0 & 1114  &  874 & -536.179025 & -536.192515 & -35.42 &   4.0 \\
a$\sigma$QZ0 & 1834  & 1580 & -536.190249 & -536.203688 & -35.28 &    41 \\
\hline
\multicolumn{7}{|c|}{MP2} \\
\hline
aDZ &  600 &  412 & -537.924103 & -537.935500 & -29.92 &   0.7 \\
aTZ & 1062 &  874 & -538.386054 & -538.396868 & -28.39 &   5.7 \\
aQZ & 1728 & 1580 & -538.533024 & -538.543616 & -27.81 &    62 \\
 \hline
a$\sigma$DZ0 &  652  &  412 & -538.117819 & -538.128461 & -27.94 &   0.6 \\  
a$\sigma$TZ0 & 1114  &  874 & -538.447985 & -538.458688 & -28.10 &   5.0 \\
a$\sigma$QZ0 & 1834  & 1580 & -538.552206 & -538.562745 & -27.67 &    52 \\
\hline
\end{tabular}
\end{center}
\end{table}

\section{\label{sec:4}Conclusions}

A new family of reduced SIGMA basis sets, ranging in size from DZ to QZ, has been reported in both standard and augmented versions, covering atoms from H to Ar. The main difference from the previous \sBS lies in the reduced number of primitive functions and the removal of the constraint that requires exponents of primitive functions appearing in a shell with angular momentum $L$ to also be present in all shells with $l < L$.
The (a)\rsXZ sets follow the previously reported matryoshka-like construction scheme. In the augmented versions, diffuse primitives with higher exponents than those in the corresponding XZ sets are added. In the $s$ and $p$ shells, these diffuse functions are incorporated into the contractions, while in the remaining shells they are included without contraction. Furthermore, unlike in the aXZ sets, the exponents of all primitive functions in the \arsXZ sets are reoptimized.
The standard versions of XZ and \rsXZ deliver comparable performance in molecular calculations. However, the augmented \arsXZ sets outperform their aXZ counterparts in terms of energy and geometry optimization, as well as computational cost.

Ancillary material is supplied in file
reduced-Sigma-BS.txt which contains the (a)\rsBS for the selected atoms in MOLPRO’s format.


\end{document}